\documentclass[11pt]{article}
\usepackage{moriond,epsfig}
\usepackage{amsmath,amssymb}

\def\be{\begin{equation}}
\def\ee{\end{equation}}
\def\bea{\begin{eqnarray}}
\def\eea{\end{eqnarray}}

\newcommand\mcN{\mathcal N}
\newcommand\mpl{m_{\rm p}}
\begin{document}
\vspace*{4cm}
\title{ INFLATION, STRINGS, CMB ANISOTROPIES AND BARYOGENESIS }

\author{ R. JEANNEROT }

\address{Instituut-Lorentz for Theoretical Physics,\\
Niels Bohrweg 2, 2333 CA Leiden, The Netherlands}

\maketitle\abstracts{In this talk, I shall focus on theories beyond
the Standard Model which predict massive neutrinos. Hybrid inflation
emerges naturally in these theories: the slow-rolling inflaton field
is a gauge singlet which couples with a GUT Higgs field which triggers
the end of inflation. In the standard scenario, spontaneous symmetry
breaking takes place at the end of inflation at a scale $M$; $M_{GUT}
> M > M_Z$ for inflation to solve the GUT monopole problem and cosmic
strings always form at this intermediate scale. WMAP data constrain $M
\in [10^{14.5}-10^{15.5}]$ GeV and the singlet-Higgs coupling $\kappa
\in [10^{-7} - 10^{-2}]$. The spectral index $n_s \gtrsim 0.98$ in
slight conflict with WMAP3. When the symmetry which is broken at
the end of inflation is gauged $B-L$, both the inflaton and the
strings decay into right-handed neutrinos. There are then two
competing non-thermal scenarios for baryogenesis via leptogenesis
which take place at the end of inflation, during reheating and from
cosmic strings decay. Which of the two scenarios dominates depends on
the inflaton-neutrino sector parameters.}
\section{Introduction}

Up to the discovery of the 'acoustic' peaks in the CMB power
spectrum~\cite{Boomerang}, there were two compelling mechanisms for
explaining cosmological perturbations: inflation and cosmic
strings~\cite{ShelVil}. Since cosmic strings predict a single peak,
they are now excluded as main source of the cosmological
perturbations. However, a mixed scenario with both inflation and cosmic
strings with a string contribution less than about 10\% is still allowed by
the data~\cite{Pogosian,WMAP}. In many models with both inflation and
strings, the scalar perturbations are dominated by the scalar
perturbations from inflation, and the string contribution may be too
low for detection via the CMB temperature anisotropies.  However they
could be detected via the B-type polarization of the
CMB~\cite{Seljak}.

From a theoretical point of view, inflation is often associated
with the formation of cosmic strings. Perhaps the best particle
physics motivated model of inflation is hybrid
inflation~\cite{hybrid}. It arises naturally in Supersymmetric (SUSY)
Grand unified Theories (GUTs)~\cite{Dvasha,prd}, in effective strings
theories and in brane worlds. Naturally meaning that the fields and
the potential leading to hybrid inflation are needed to build the
theory itself (I now focus on the case of SUSY GUTs) and the coupling
constant which enters is the order unity.  In either cases,
spontaneous symmetry breaking takes place at the end of
inflation~\footnote{In non minimal models of hybrid inflation such as
shifted inflation~\cite{shifted} or smooth inflation~\cite{smooth},
spontaneous symmetry breaking takes place before or during inflation,
and no defect form at the of inflation.}  and cosmic strings
form~\cite{prd,jrs,csform,Dstrings}. In this talk I will consider
Standard hybrid inflation in the context of SUSY GUTs. And I shall be mainly concerned about models which contain $B-L$ as a gauge symmetry and predict
massive neutrinos via the See-saw mechanism~\cite{seesaw}.

In section~\ref{sec-cs}, I show that cosmic strings always form at the
end of standard~\footnote{{\it Standard} refers to the standard model
of SUSY hybrid inflation~\cite{Dvasha} where SSB takes place at the
end of inflation} hybrid inflation when inflation solves the GUT
monopole problem. In section~\ref{sec-cmb}, I study the CMB
anisotropies which are predicted by these models. Matching theoretical
predictions with the data gives constraints on two of the GUT
parameters, the Spontaneous Symmetry Breaking (SSB) scale at the end
of inflation and the relevant coupling constant~\cite{cmb,infl}. In
section~\ref{sec-lepto}, I show that when the symmetry which is
broken at the end of inflation is gauged $B-L$, there are two
competing non-thermal baryogenesis scenarios which take place after
inflation: from reheating, and from cosmic strings decay~\cite{lept}.

\section{Inflation and cosmic strings}
\label{sec-cs}

\subsection{Inflation from particle physics}

Inflation must come from the particle physics model describing
fundamental interactions at high energies. As a particle physicist, the
first question i will ask is 'Can we get inflation from the Standard
Model?' On general grounds, the answer is 'No', because the
inflationary energy scale would be the order of $100$ GeV which is far
too low to produced the required amount of primordial
perturbations~\footnote{However it has been recently suggested that an
MSSM flat direction might be suitable for inflation
~\cite{MSSMflat}. Even though this proposal requires strong
fine-tuning, it is interesting in two ways: first of all it uses
standard model physics, and second there is no need of standard model
singlet.}. The next question i will ask is 'Can we get inflation from
the simplest extensions of the Standard Model?'  As an apart\'e, we
know since the discovery of neutrinos oscillations that neutrinos are
massive and hence that the Standard Model must be extended. In order
to explain the smallness of the observed neutrinos masses, one could
just add a gauge singlet and a tiny coupling constant. However, by
adding a $U(1)_{B-L}$ gauge symmetry to the Standard Model gauge group
$G_{SM} = SU(3)_c \times SU(2)_L \times U(1)_Y$, massive neutrinos
become a prediction~\cite{seesaw}. Adding the idea of unification of
the gauge coupling constants, one is lead to grand unified theories.
So I shall rephrase the question as 'Can we get inflation from a grand
unified theory?' At first sight, 'the unification scale $M_{GUT}
\sim 10^{16}$ GeV is just the energy scale needed for inflation to
explain the cosmological perturbations'.

It turns out that when building a model of slow-roll inflation in a
theory beyond the Standard Model three ingredients are usually needed:
SUSY, which provides the required flatness of the potential, a
Standard Model singlet, the slow-rolling field, and GUT Higgs fields
transforming under a gauge group $G$ whose rank is larger
than the rank of the Standard Model gauge group, i.e. rank$(G)> 4$
~\cite{prd,csform}.

\subsection{Standard hybrid inflation}

Hybrid inflation~\cite{hybrid} uses two fields instead of one, a gauge
singlet $S$ and a Higgs field $\Phi$. Hybrid inflation is arguably the
best particle physics motivated model of inflation. In the context of
SUSY GUTs, there are two Higgs superfields $\Phi$ and $\bar{\Phi}$ in
complex conjugate representations of the GUT gauge group $G_{GUT}$
which lower the rank of the group by one unit when acquiring
vacuum expectation values (VEV) at the end of inflation.  The
superpotential is given by
\begin{equation} 
W_{\rm inf} = \kappa S( \bar{\Phi} \Phi - M^2),
\label{eq:W}
\end{equation}
where a suitable U(1) R-symmetry under which $W$ and $S$ transform in
the same way ensures the uniqueness of this superpotential at the
renormalizable level.  The scalar potential has an inflationary
valley, which is a valley of local minima, at $S > M$ and $|\Phi| =
|\bar{\Phi}| = 0$. At tree level, the potential along this valley is
$V_{infl} = \kappa^2 M^2$. $S$ is the slowing rolling field and
slow-roll conditions thus apply to $S$. Since $|\Phi| = |\bar{\Phi}| =
0$ during inflation, there is no symmetry breaking induced by these
Higgs fields VEV during inflation. Inflation terminates as $S$ falls
below its critical value $S_c = M$ and inflation ends in a phase
transition during which the Higgs fields acquire non-zero VEV equal to
$M$: Spontaneous Symmetry Breaking (SSB) takes place at the end of
inflation. The SSB scale $M$ and is proportional to the inflationary
scale $V_{infl}^{1/4}$, the proportionality coefficient being the
squared root of the singlet-Higgs coupling $\kappa$.

The Higgs fields representations $\Phi$ and $\bar{\Phi}$ are conjugate
N-dimensional representations of the GUT gauge group. We are now
focusing on GUT which contain gauged $U(1)_{B-L}$ and predict massive
neutrinos via See-saw. The component of of $\Phi$ (and $\bar{\Phi}$)
which gets a VEV at the end of inflation transforms as an Standard
Model singlet, and it also transforms either as an $SU(2)_R$ doublet
or as an $SU(2)_R$ triplet. In a realistic model where there are no
unwanted light fields between the scale $M$ and the GUT scale, it
is the only component which remains light below $M_{GUT}$~\cite{infl}
($M < M_{GUT}$, see section 2.3). The scalar potential along the
inflationary valley is flat at tree level. It is lifted by loop
corrections, which are non-zero during inflation because SUSY is
spontaneously broken, and by SUGRA corrections. Assuming minimum
Kh\"aler potential it is given by ~\cite{cmb,infl}
\begin{eqnarray} 
\frac{V}{\kappa^2 M^4} &=& 1 +
\frac{\kappa^2 \mcN}{32 \pi^2}
\Big[ 2 \ln \Big(\frac{\kappa^2 M^2 x^2}{\Lambda^2}\Big) + (x^2+1)^2
\ln(1+x^{-2}) + (x^2-1)^2 \ln (1-x^{-2}) \Big] 
\nonumber \\
&+& 
2 x^4 \Big( \frac{M}{\mpl}\Big)^4 + |a|^2 x^2 \Big( \frac{M}{\mpl}\Big)^2
+ A \frac{m_{3/2}}{M} x,
\label{eq:V}
\end{eqnarray}
where $\mpl$ is the reduced Planck mass and $\Lambda$ a cutoff scale; $x
= |S|/M$ so that $x\to 1$ at the critical point; $A= 4 \cos(\arg
m_{3/2} - \arg S)$, we assume that $\arg S$ is constant
during inflation; $\mcN=1-3$ depending on wether the components of
$\Phi$ and $\bar{\Phi}$ which get a VEV at the end of inflation
transform as an $SU(2)_R$ doublet or triplet and wether the symmetry
group which breaks at the end of inflation contains an $SU(2)_R$ or an
$U(1)_R$ symmetry. Hidden sector VEV which lead to low energy SUSY
breaking are $\langle z \rangle = a \mpl$ and $\quad \langle W_{\rm hid}(z)
\rangle = m_{3/2} \exp^{- |a|^2/2} \mpl^2$, with $m_{3/2}$ the
gravitino mass; the cosmological constant in the global minimum is set
to zero by hand. All subdominant terms are dropped.

\subsection{Cosmic strings form at the end of standard hybrid inflation}

Since SSB takes place at the end of inflation, cosmic strings always
form if the later solves the GUT monopole problem~\cite{csform}. The
underlying reason being that the rank of the gauge group is lowered by
one unit at the end of inflation~\cite{prd,csform}. This is
illustrated in reference~\cite{jrs} where an exhaustive study of all
SSB breaking patterns for all GUT gauge groups with rank less than
height and phenomenologically acceptable has been performed. The aim of
this section is to understand why indeed cosmic strings form. Further
details can fe found in reference~\cite{csform}.

Suppose that the Standard Model gauge group $G_{SM}$ is embedded in a
GUT gauge group $G_{GUT}$. This must be broken down to $G_{SM}$ at
around $M_{GUT} \sim 10^{16}$ GeV, which is the scale at which the
gauge couplings unify
\begin{equation}
G_{GUT} \stackrel{M_{GUT}}\rightarrow ...\rightarrow
G_{SM} \stackrel{10^2 \,\,\rm GeV}\rightarrow SU(3)_c \times U(1)_Q.
\end{equation}
In SUSY, the breaking of $G_{GUT}$ down to $G_{SM}$ can be direct of
via intermediate symmetry groups, whereas in the non SUSY case there
must be at least one intermediate step. If (some of) the Higgs fields
used to break $G_{GUT}$ have a superpotential given by
Eq.(\ref{eq:W})~\footnote{The rank of $G_{GUT}$ has to be strictly
greater than the rank of $G_{SM}$~\cite{csform}}, inflation takes
place and the spontaneous symmetry breaking of $G_{GUT}$ takes place at
the end of inflation. But in this scenario, cosmologically
catastrophic monopoles which ought to form in all GUTs, form after
inflation. In order to cure the monopole problem, one must introduce
an intermediate symmetry group $H$, a subgroup of $G_{GUT}$, and use
$\Phi$, $\bar{\Phi}$ not to break the GUT itself but this intermediate
symmetry group $H$; the symmetry breaking scale $M$ of $H$ is
$<M_{GUT}$. $H$ must be chosen in such a way that the monopoles form
between $G_{GUT}$ and $H$ and no unwanted defect form when $H$ breaks
down to $G_{SM}$
\begin{equation}
G_{GUT}\stackrel{\rm Monopoles}\rightarrow...H \stackrel{\Phi, \bar{\Phi}, \,\, \rm No\,\, unwanted \,\, defect}\rightarrow ...\rightarrow G_{SM}
\stackrel{10^2 \,\,\rm GeV}\rightarrow SU(3)_c \times U(1)_Q.
\end{equation}
It can be shown that the rank of $H$ must be greater than five,
that it must contain a U(1) factor~\cite{prd,csform}, and that cosmic strings
always form when $H$ breaks down to $G_{SM}$~\cite{prd,jrs,csform}.
\begin{equation}
G_{GUT}\stackrel{\rm Monopoles}\rightarrow...H \stackrel{\rm Inflation, \,\,\rm  Cosmic \; Strings}\rightarrow ...\rightarrow G_{SM}
\stackrel{10^2 \,\,\rm GeV}\rightarrow SU(3)_c \times U(1)_Q.
\end{equation}

\section{CMB constraints and predictions}
\label{sec-cmb}

If the strings which form at the end of inflation are stable down to
low energy, they will contribute to the CMB temperature
anisotropies. The perturbations from inflation and cosmic strings are
uncorrelated and they add up independently~\cite{prd}

\begin{equation}
\Big(
\frac{\delta T }{T} \Big)_{\rm tot} = \sqrt{ \Big( \frac{\delta T }{T} \Big)_{\rm inf}^2
+ \Big( \frac{\delta T }{T} \Big)_{\rm cs}^2 }.
\label{dTtot}
\end{equation}

The inflation contribution to the quadrapole is
\begin{equation}
\Big( \frac{\delta T }{T} \Big)_{\rm inf}
= \frac{1}{12\sqrt{5} \pi \mpl^3} \frac{V^{3/2}}{V'} 
\Bigg|_{\sigma = \sigma_Q},
\label{dTphi}
\end{equation}
with a prime denoting derivative w.r.t. the real normalized inflaton
field $\sigma = \sqrt{2} |S|$, and the subscript $Q$ denoting the time
observable scales leave the horizon. $V$ is the scalar potential along
the inflationary trajectory given by Eq.(\ref{eq:V}). The tensor
perturbations from inflation $(\delta T/T)_{\rm tens} \sim 10^{-2}
H/\mpl$ are very small.

The string induced perturbations are proportional to the string
tension $(\delta T/T)_{\rm cs} = y G \mu$, with $\mu$ the tension and
$y$ parameterizing the density of the string network. Recent
simulations predicts $y=9 \pm 2.5$ \cite{Landriau}, but values in the
range $y=3-12$ can be found in the literature \cite{ShelVil}. The
strings are formed by the Higgs fields $\Phi$ and $\bar{\Phi}$ which
wind around the string in opposite directions. They are not BPS and do
not satisfy the Bogomolnyi bound and hence they are lighter than BPS
strings forming at the same energy scale \cite{cmb}. The string
tension is
\begin{equation} 
\mu = 2 \pi M^2 \theta(\beta),
\quad  {\rm with} \quad
\theta(\beta) = 
\sim 2.4 \ln (2/\beta)^{-1} 
\end{equation}
where the function $\theta$ encodes the correction away from the BPS
limit and $\beta = (m_\phi / m_A)^2 \simeq (\kappa/g_{\rm GUT})^2$
with $g_{\rm GUT}^2 \approx 4\pi/25$.  Requiring the
non-adiabatic string contribution to the quadrupole to be less than
10\% gives the bound \cite{cmb}
\begin{equation}
G \mu < 2.3 \times 10^{-7} \Big( \frac{9}{y} \Big) 
\quad \Rightarrow \quad
M < 2.3 \times 10^{15} 
\sqrt{\frac{(9/y)}{\theta(\beta)}}.
\label{Pogosian}
\end{equation}
The bound which comes from pulsar timing (the stochastic gravitational
wave background produced by cosmic strings can disrupt pulsar timing
and this has not been observed) is $G \mu < 1.0 \times
10^{-7}$~\cite{Lommen}; it is more stringent, but it has also more
uncertainties. It corresponds to the 10\% bound with $y=20.7$.

\begin{figure}[h]
\begin{center}
\leavevmode\epsfysize=5.5cm \epsfbox{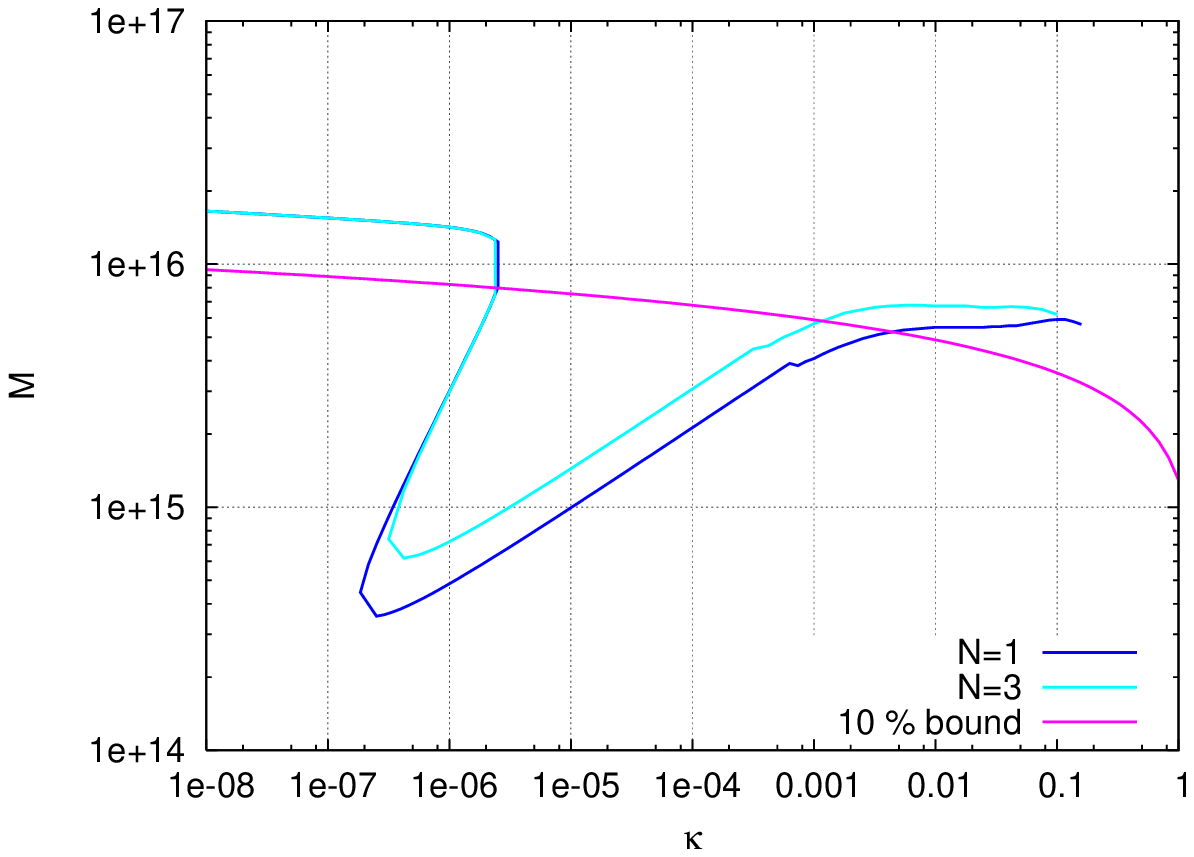}
\leavevmode\epsfysize=5.5cm \epsfbox{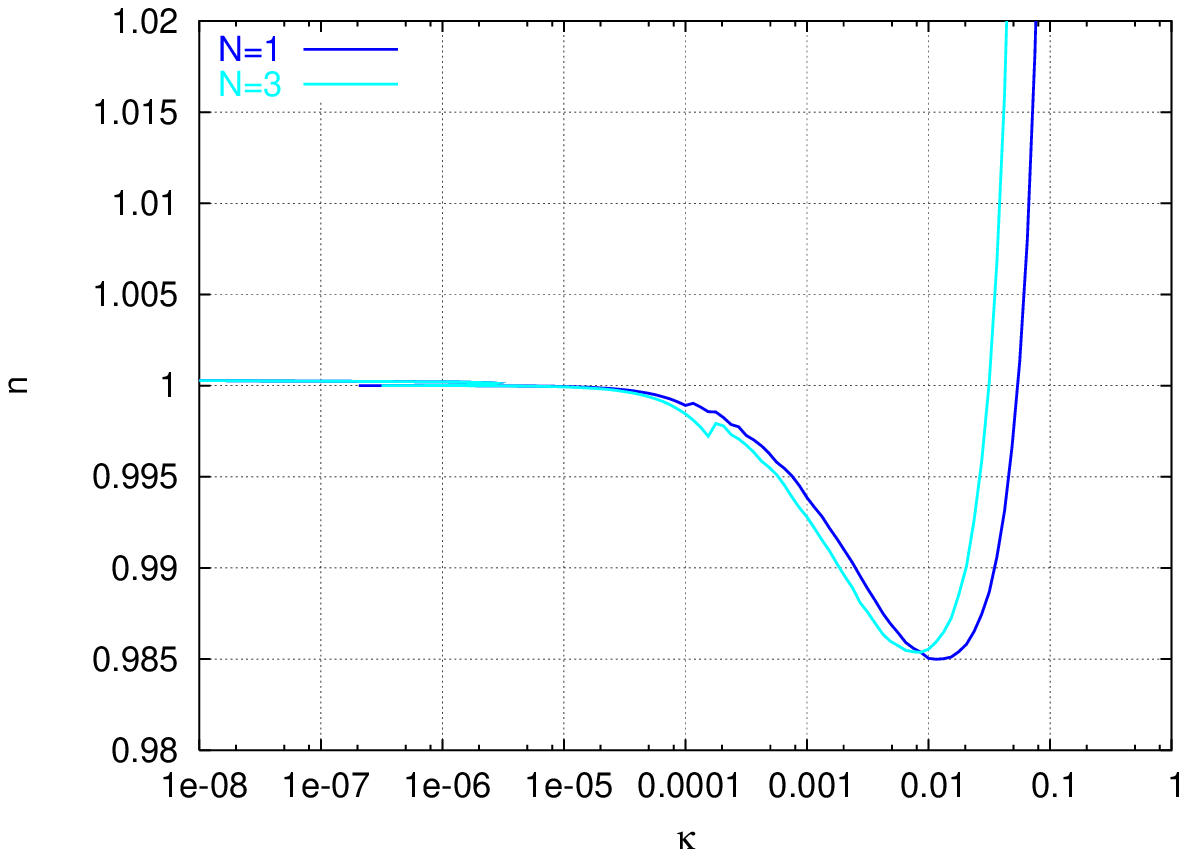}
\caption{Left: CMB constraints on $M$ as a function $\kappa$ for
$\mcN=1,3$ (blue curves) and the 10\% bound (pink curve). Right:
predictions for the spectral index $n_s$ as function of $\kappa$ for $\mcN=1,3$.}
\label{F:F}
\end{center}
\end{figure}

Temperature anisotropies from both inflation and cosmic strings depend
on two parameters, the SSB scale $M$ of the intermediate symmetry
group $H$, see Sec 2., and the singlet-Higgs coupling constant
$\kappa$, see Eq.(\ref{eq:W}). Matching the theoretical predictions
with the observed value $(\delta T/T) = 6.6 \times 10^{-6}$
\cite{WMAP} gives a constrain on $M$ versus $\kappa$, see
figure~\ref{F:F} \cite{cmb,infl}. The intermediate symmetry breaking
scale must be very close to the GUT scale, $M\in [10^{14.5} -
10^{15.5}]$ GeV and the coupling constant $\kappa \in
[10^{-7}-10^{-2}]$. If the strings are unstable~\cite{infl}, larger
values of $\kappa$ are allowed.

The spectral index $n_s$ is calculated using the slow-roll parameters
and also depends on the singlet-Higgs coupling constant $\kappa$; it
is also shown on figure~\ref{F:F}~\cite{cmb,infl}. $n_s$ is
undistinguishable from unity for small values of $\kappa$, smaller
than unity for intermediate values of $\kappa$ and bigger than unity
for large values of $\kappa$. It is extremely difficult to get
a spectral index smaller than $0.98$ with hybrid inflation except maybe
with non-minimal models \cite{Shafi}. This is in slight conflict with
WMAP 3-years data, and if these were to be confirmed, it would be
excluded.

\begin{figure}[h]
\begin{center}
\leavevmode\epsfysize=5.5cm \epsfbox{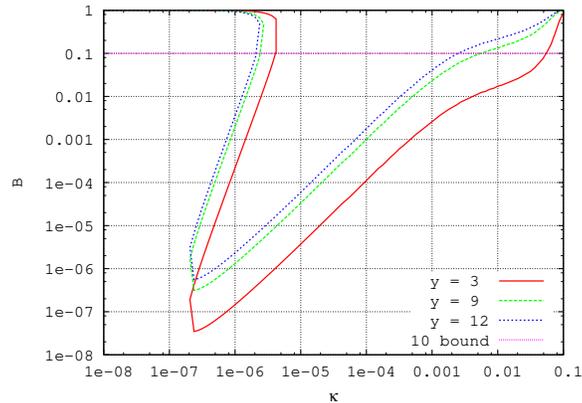}
\caption{The string contribution $\Big(\frac{\delta T }{T} \Big)_{tot}/\Big( \frac{\delta T }{T} \Big)_{\rm cs}$ as a function of $\kappa$ for $y = 3, 6, 9$.}
\label{F:B}
\end{center}
\end{figure}

The string contribution $B = \Big(\frac{\delta T }{T} \Big)_{\rm
tot}/\Big( \frac{\delta T }{T} \Big)_{\rm cs}$ is also a function of
the coupling $\kappa$. It is negligibly small for most of the
parameter space and saturates the 10\% bound for large values of
$\kappa$, which is the best interesting region for both $n_s$ and
baryogenesis (see Sec. 4). It is shown in figure~\ref{F:B}~\cite{cmb}. 

Further details can be found in references~\cite{cmb,infl}.

\section{Baryogenesis via leptogenesis at the end of inflation}
\label{sec-lepto}

Baryogenesis aims to explain the observed matter-antimatter asymmetry
of the Universe. It must take place after inflation, since any
previously produced baryon asymmetry is washed-out. Standard GUT
baryogenesis is ruled out because any GUT scale produced baryon
asymmetry is erased by sphalerons transitions unless the
universe possesses a $B-L$ asymmetry~\footnote{Sphalerons transition
violate $B+L$ and conserve $B-L$, where $B$ and $L$ are respectively
number and lepton number.}~\cite{Rubakov}. A primordial $B-L$
asymmetry is naturally obtained in theories beyond the standard model
which contain gauged $B-L$ via the out-of-equilibrium
decay of heavy Majorana right-handed neutrinos~\cite{Yanagida}. This
scenario known has leptogenesis is perhaps the best particle physics
motivated model of baryogenesis. Thermal leptogenesis requires a
symmetry breaking scale $\sim 10^{15}$ GeV and a reheating temperature
$T_R \sim 10^{10}$ GeV~\cite{Buchmuller}. Such high reheating
temperature leads to an overproduction of gravitinos which decay
lately and disrupt the predictions of nucleosynthesis. 

When $G_{GUT}, H \supset U(1)_{B-L}$ and the $\Phi$ and $\bar{\Phi}$
fields entering the inflation superpotential given by equation (\ref{eq:W})
are the $B-L$ breaking Higgs fields, gauged $B-L$ is broken at the end
of inflation and the strings which form at the end of inflation are
the so-called $B-L$ cosmic strings~\cite{prl}. There are then two
competing non-thermal scenarios for leptogenesis which take place
after inflation: from reheating during inflation \cite{lepto1} and
from cosmic strings decay \cite{lept,prl}.

\begin{itemize}
\item Non-thermal leptogenesis during reheating

The $B-L$ breaking Higgs field $\Phi$ which enters the inflationary
superpotential Eq.(\ref{eq:W}) gives a superheavy Majorana mass to the
right-handed neutrinos ($W\supset \Phi NN$ or $W \supset
\Phi^2NN/\mpl$) and reheating proceeds via production of heavy
right-handed neutrinos and sneutrinos. Right-handed (s)neutrinos decay
into electroweak Higgs(ino) and (s)leptons ($W \supset H_u L N$), CP
is violated through the one-loop radiative correction involving a
Higgs particle and by the self-energy correction, and lepton asymmetry
is non-thermally produced when the right-handed neutrinos are
out-of-equilibrium, i.e. when $T_R < M_{N_i}$. If $T_R>M_{N_1}$, the
lepton asymmetry produced is wash-out by L-violating processes
involving right-handed neutrinos until $T <M_{N_1}$, where $M_{N_1}$
is the mass of the lightest right-handed neutrino.  If $M_{N_1} >
m_\Phi/2$, where $m_{\Phi}$ is the mass of the Higgs field in the true
vacuum, the inflaton cannot decay into right-handed neutrinos and
reheating must be gravitational. The resulting baryon asymmetry
depends on the reheating temperature at the end of inflation, which
depends on the mass $M_{N_i}$ of the heaviest right-handed neutrinos
the inflaton can decay into, on the symmetry breaking scale $M$
which is constrained by CMB data as a function of the coupling
$\kappa$ (see Sec. 3) and on the CP violating parameter~\cite{lept}.

\item Non-thermal leptogenesis from cosmic strings decay

The strings which form at the end of inflation are the so-called $B-L$
cosmic strings \cite{prl}. The main decay channel of $B-L$ strings is
into right-handed neutrinos and they also lead to non-thermal
leptogenesis \cite{prl,lept}. The resulting baryon asymmetry depends upon the
amount of energy loss by the network into right-handed neutrinos and
on the density of strings at the end of inflation; it also depends on
the symmetry breaking scale $M$ at the end of inflation which is
constrain by CMB as a function of $\kappa$~\cite{lept}.
\end{itemize}
Which of these two scenarios dominates depends on wether the inflaton
decay into right-handed neutrino is kinematically allowed and wether
the lightest right-handed neutrino $N_1$ is in thermal equilibrium at
reheating. Results, which take into account the CMB constraints
derived in the previous section, are shown in figures 3 and 4. Further
details can be found in reference~\cite{lept}.
\begin{figure}[h]
\begin{center}
\leavevmode\epsfysize=5.5cm \epsfbox{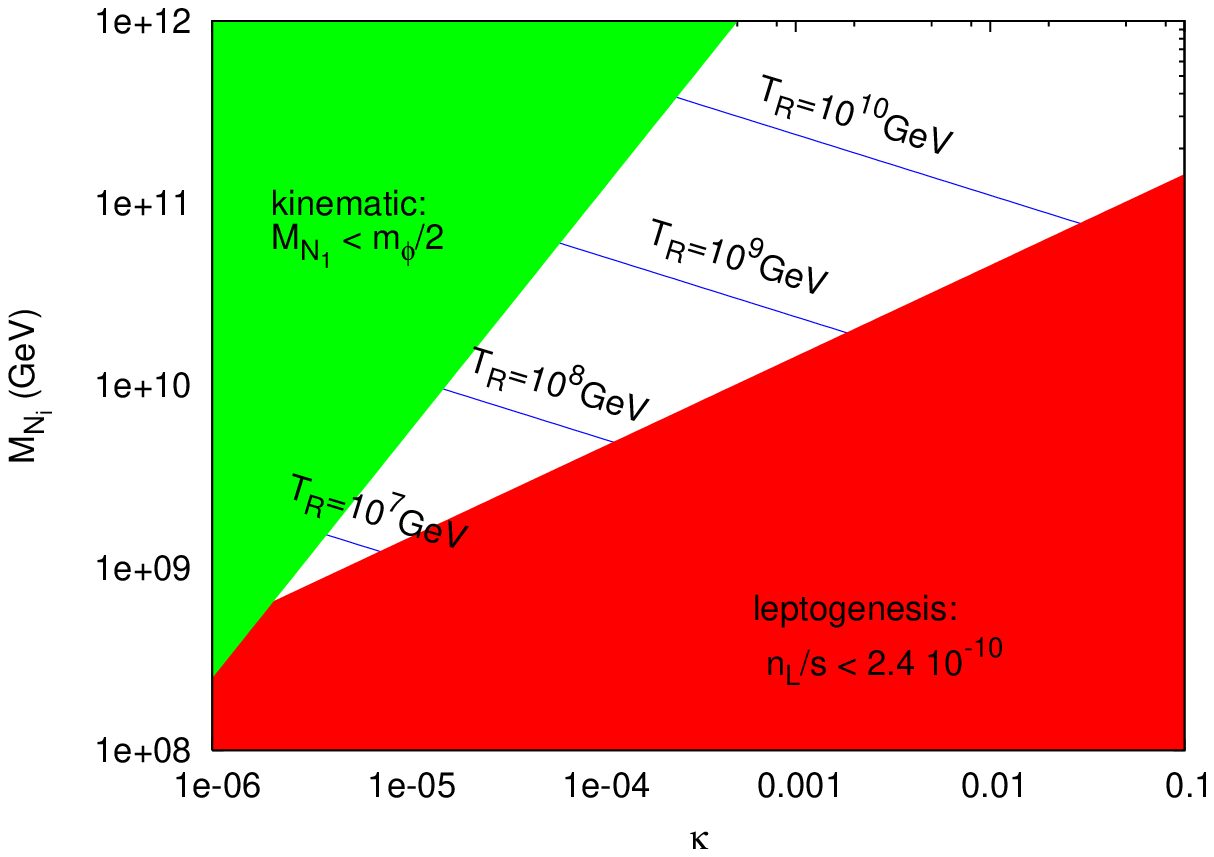}

\leavevmode\epsfysize=5.5cm \epsfbox{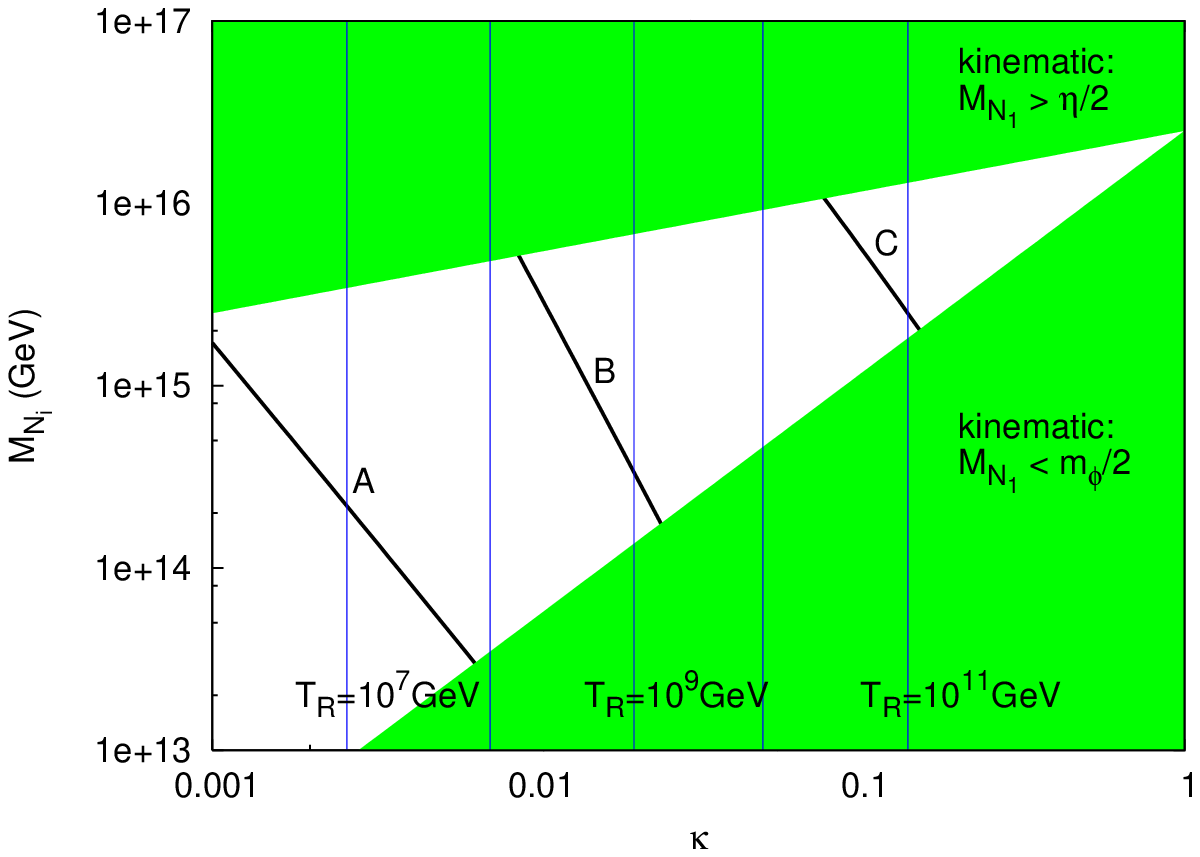}
\leavevmode\epsfysize=5.5cm \epsfbox{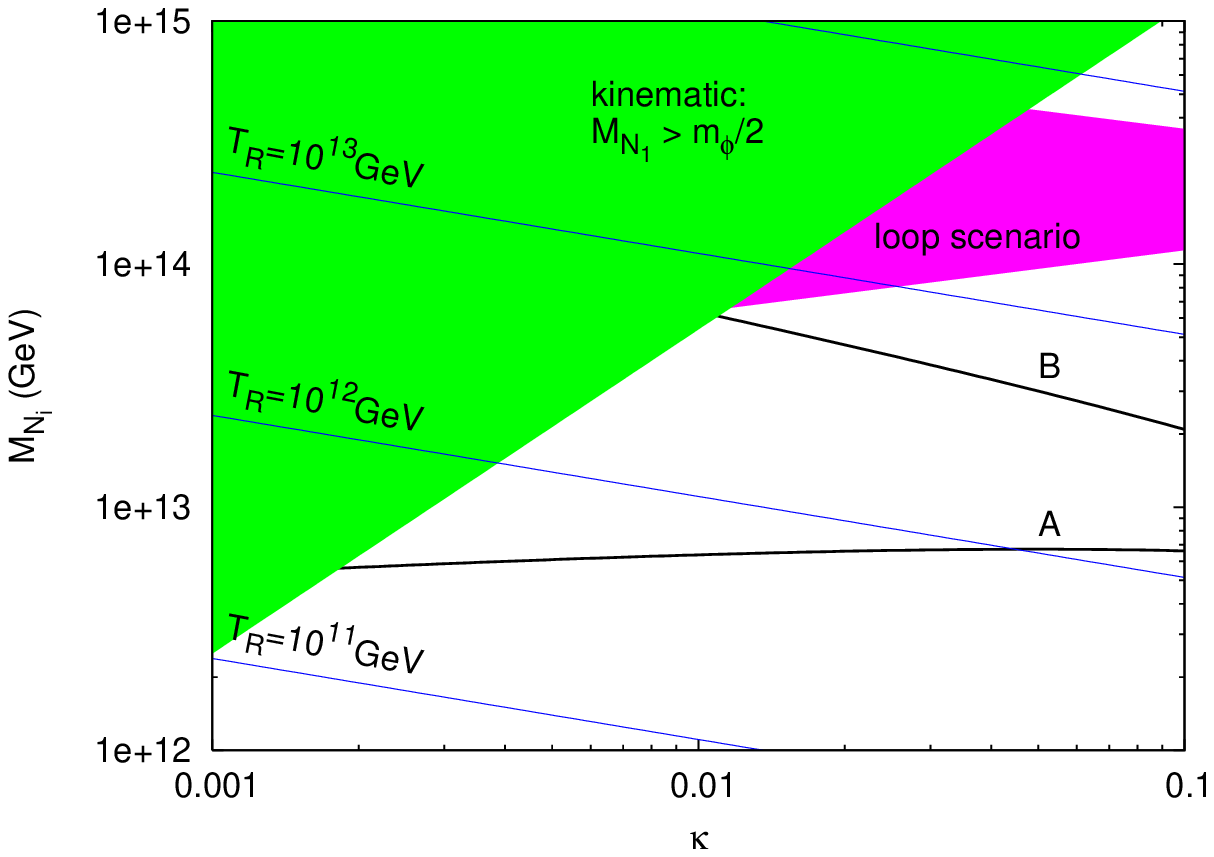}
\caption{The mass $M_{N_i}$ of the heaviest right-handed neutrino the
inflaton can decay into as a function of the coupling $\kappa$. The
white regions give the measured value of $n_B/s$. The colored regions
are excluded. Top: $M_{N_1} < m_\phi/2$ and $M_{N_i} > T_R$, both
strings and inflation contribute non-thermally to $\Delta B$. Bottom:
$M_{N_1} > m_\phi/2$, reheating is gravitational and only strings
contribute non-thermally to $\Delta B$. Bottom Left: $M_{N_1} >T_R$, there is
no wash out. Bottom Right: $M_{N_1} < T_R$, there is also a thermal
contribution.}
\end{center}
\end{figure}

\section{Conclusions}

GUT which predict massive neutrinos are good candidates for hybrid
inflation. Cosmic strings form at the end of inflation and, if they
stable down to low energy, they contribute to CMB anisotropies
together with inflation. The symmetry breaking scale is constrained by
CMB data to the range $M \in [10^{14.5}-10^{15.5}]$ GeV and the
relevant coupling $\kappa \in [10^{-7} - 10^{-2}]$.  Scalar
perturbations from inflation dominate for a large part of the
parameter space and it might be impossible to detect the strings using
the temperature anisotropies of the CMB.  They could however be
detected via the B-type polarization of the CMB \cite{Seljak}.

Hybrid inflation predicts a spectral index $n_s \gtrsim 0.98$. It is
very difficult to get smaller values except maybe by going to non
minimal models \cite{Shafi}; hence if the three year WMAP central
value $n_s = 0.951^{+0.015}_{-0.019}$ were to be confirmed, hybrid
inflation with minimal SUGRA could be excluded. But even if scalar
perturbations are dominated by scalar perturbations from inflation,
tensor perturbations (which are negligible for hybrid inflation) can
nonetheless be dominated by tensor perturbations from cosmic
strings; this can allow a larger value of $n_s$~\cite{WMAP}.

Finally, when $B-L$ is broken at the end of inflation, baryogenesis
via leptogenesis takes place after inflation during reheating and/or via cosmic strings decay; which of the two scenarios dominates depends upon the 
various parameters in the inflaton-neutrino sector.

\section*{Acknowledgments}

I wish to acknowledge Marieke Postma for enjoyable collaboration. I
also wish to thank The Dutch Organization for Scientific Research
[NWO] for financial support.

\end{document}